\documentclass[useAMS,referee]{biom}
\usepackage{amsmath}
\usepackage{amsfonts}
\usepackage{graphicx}
\usepackage[colorlinks=true, allcolors=blue]{hyperref}
\usepackage{float}
\usepackage{multicol}
\usepackage{dsfont}

\def\bSig\mathbf{\Sigma}

\newcommand{\bepsilon}{\mbox{\boldmath $\epsilon$}}

\newcommand{\bbeta}{\mbox{\boldmath $\beta$}}
\newcommand{\btheta}{\mbox{\boldmath $\theta$}}

\newcommand{\bSigma}{\mbox{\boldmath $\Sigma$}}

\newcommand{\bgamma}{\mbox{\boldmath $\gamma$}}

\newcommand{\Y}{\mathbf{Y}}

\newcommand{\Z}{\mathbf{Z}}

\newcommand{\A}{\mathbf{A}}

\newcommand{\X}{\mathbf{X}}
\newcommand{\W}{\mathbf{W}}

\newcommand{\w}{\mathbf{w}}

\newcommand{\one}{\mathbf{1}}

\newcommand{\bpi}{\mbox{\boldmath $\pi$}}

\newcommand{\Var}{{\mbox{Var}}}
\newcommand{\Cov}{{\mbox{Cov}}}
\newcommand{\undertilde}[1]{\underset{\widetilde{}}{#1}}

\title[Corrections in Linear Regression]{Correction of estimator bias in linear regression with
categorical covariates with classification error}

\author{Alexandre Garcia Dias$^{*}$\email{a163386@dac.unicamp.br},
Mariana Rodrigues Motta$^{**}$\email{marirm@unicamp.br}\\
Department of Statistics, University of Campinas, Campinas, Brazil
\and
Alexandre Hild Aono \\
Center for Molecular Biology and Genetic Engineering (CBMEG), \\
University of Campinas (UNICAMP), Campinas, Brazil}

\begin{document}


\date{{\it Received October} 2007. {\it Revised February} 2008.  {\it
Accepted March} 2008.}



\pagerange{\pageref{firstpage}--\pageref{lastpage}} 
\volume{64}
\pubyear{2008}
\artmonth{December}


\doi{10.1111/j.1541-0420.2005.00454.x}


\label{firstpage}


\begin{abstract}
The objective of this work is to propose an asymptotic correction method for the estimators
of parameters from regression models with covariates subject to classification errors. A
correction was developed based on the least squares estimators from regression with
erroneous covariates, the marginal probability of the true covariates, and the conditional
probability of the erroneous covariates given the true covariates. In this way, we can correct
these estimators without the need to correct the erroneous covariates or observe the true
covariates. We performed simulations to quantify the performance of the proposed
corrections, identifying, that correcting the intercept is crucial for a significant
improvement in estimation.
\end{abstract}

%



\maketitle


%

\section{Introduction}
\label{s:intro}

A major limitation in traditional linear models is that they typically treat covariates as fixed and error-free. This assumption is problematic, especially in medical and biological research, where covariate measurement errors are often underreported and rarely corrected \cite{brakenhoff2018}. Despite the availability of various correction techniques for continuous covariates—such as moment-based adjustments, quasi-likelihood calibration, and SIMEX—the ordinary least squares (OLS) method without corrections remains widely used, potentially leading to biased and inconsistent estimates.

For categorical covariates, error correction methods are less developed due to assumptions (e.g., normal additive error) that do not hold. Some existing approaches include data augmentation \cite{kuha}, nonparametric estimation \cite{chen}, and score-based correction \cite{zucker}. Buonaccorsi, in his 2005 article, proposed a bias correction method for binary covariates using the covariance matrix between observed and true variables.

The main focus of the referenced work is to extend Buonaccorsi’s method to handle multiple multinomial covariates, where each covariate may have different category levels. This extension is
particularly motivated by quantitative genomics, a field that uses genome sequencing data to study
the genetic basis of complex traits (CITE). In this context, researchers identify genetic mutations by
comparing differences in DNA sequences across many individuals in a population. A genomic
segment can be defined as a sequence of nucleotides, represented by the letters A (adenine), T
(thymine), C (cytosine), or G (guanine).

A polymorphism refers to a position in the genome where variation exists between individuals in
a population. The most common type of polymorphism used in genetic studies is the single
nucleotide polymorphism (SNP), where the variation affects only a single nucleotide position
(CITE). For example, at a particular position in the genome, some individuals might have the
nucleotide A, while others have T. These differences represent alternative versions of a DNA
sequence, known as alleles.

The way SNPs are encoded depends on the species’ ploidy, which is the number of complete sets
of chromosomes in its genome. In diploid species (such as humans), individuals inherit two sets of
chromosomes (one from each parent). In this case, each SNP position can be categorized into three possible genotypes: homozygous reference (e.g., AA), heterozygous (e.g., AT), or homozygous
alternative (e.g., TT). Because most SNPs are biallelic (they involve only two possible alleles), these
three categories are usually sufficient. Genotypes can be numerically encoded as: 0 for homozygous
reference, 1 for heterozygous, and 2 for homozygous alternative.

In species with more complex genomes, known as polyploid species, there are more than two sets
of chromosomes. For instance, tetraploid species have four sets. In such cases, SNPs can exhibit a
range of allele dosages, referring to how many copies of a specific allele are present at a given locus.
For example, a biallelic SNP with alleles A and T in a tetraploid organism could be represented as: 0
= AAAA (no copies of T), 1 = TAAA (one copy of T), 2 = TTAA (two copies of T), 3 = TTTA (three
copies of T), or 4 = TTTT (four copies of T).

This fine-scale dosage information makes genotype calling (the process of determining which
genotype an individual has) more challenging in polyploid species. The difficulty arises due to
sequencing errors, where each nucleotide read has a probability of being incorrect (CITE).
Consequently, the estimated allele dosage may be uncertain. These errors can lead to incorrect SNP
rankings and underestimation of correlations between markers and genetic traits (Hackett et
al.,2003; Go¨ring et al.,2000). Moreover, genotyping errors can greatly impact genetic studies,
reducing their efficiency and potentially leading to false conclusions in analyses such as kinship
estimation (Ward et al.,2021)

Classification problems in categorical covariates also appear in other areas of quantitative genetics. For instance, in linkage analysis—a field aimed at identifying loci responsible for specific phenotypes—two-locus methods are more robust to genotyping errors, while multi-locus methods may erroneously exclude true disease-gene loci due to misclassification \cite{goring}. Furthermore, low-density SNP panels can be imputed to high-density panels, mitigating data loss, though the accuracy of this process depends on the size of the reference population \cite{dassoneville}.

To minimize genotyping errors, \cite{ward} recommend optimizing experimental methods, using appropriate controls and replicates, and developing statistical techniques for error detection. Ideally, such practices would only lead to the exclusion of noisy or uninformative data. The challenge lies in minimizing data exclusion without distorting the information contained in error-prone data. This study proposes an asymptotic bias correction method for the estimators of a linear regression model when covariates have a substantial level of uncertainty, rather than being perfectly reliable.

\section{Model}
\label{s:model}
This section defines the linear model which has only K categorical covariates, where the $k$th covariate has $L_k$ levels, $k=1,\ldots,K$.
\subsection{Linear regression model with $K=1$ categorical variable with two or more categories which may have classification error}
\label{sing}

Consider a random variable X whose values are sit in $\{1,2,\ldots,L\}$ where $p_l=P(X=l)$, $l=1,\ldots,L \mbox{ and } \displaystyle\sum_{l=1}^{L}{p_l} =1$. Suppose that $X$ is subject to classification error and the random variable $W$ represents the observed values. Assume the categories of W are the same as those of $X$. Following (Buonacorrsi), let
\begin{eqnarray}
\theta_{l|m} &=& P(W = l|X = m), 
\label{eq:theta}
\end{eqnarray}
such that
\begin{eqnarray}
\pi_{m|l} &=& P(X = m|W = l)   \nonumber \\
            &=& \theta_{l|m} \frac{P(X=m)}{\displaystyle \sum_{m'=1}^L\theta_{l|m'}P(X=m')} \label{eq:pi_i}.
\end{eqnarray}

Conditioning on $X=m$, $m=1,\ldots,L$, $W$ has a Multinomial distribution with probability $\theta_{1|m},\ldots,\theta_{L|m}$ for the categories 1 to $L$, respectively.

Define $X_1, \ldots, X_n$ as independent random variables with the same distribution as $X$ and $W_1, \ldots, W_n$ the respective vectors with classification error. Furthermore, suppose that instead of observing $X_i$, we observe $W_i$. The auxiliary vectors $\X_i$ e $\W_i$, $i=1,\ldots,n$ are construed in such a way that their components, $X_{il}$ and $W_{il}$ are binary and defined by the following relationship: 
$X_{il}=1 \iff X_i=l$ e $W_{il}=1 \iff W_i=l$ where $l=1,2,\ldots,L-1 $. The linear model containing the variables without error is given by
\begin{eqnarray}
Y_i &=& \beta_{0} + \sum_{l=1}^{L-1} \beta_{l} X_{il} + \epsilon_{Xi}. \label{eq1:modelox}
\end{eqnarray} In matrix form we have
\begin{eqnarray}
    \Y=\mathds{1} \beta_0 + \X\bbeta + \bepsilon, 
\end{eqnarray}
where
\begin{eqnarray}
\Y&=&\begin{bmatrix}
    Y_1\\
    \vdots\\
    Y_n
\end{bmatrix},\nonumber
    \X = \begin{bmatrix}
    \X_1\\
    \vdots\\
    \X_n
\end{bmatrix},\bepsilon = \begin{bmatrix}
    \epsilon_1\\
    \vdots\\
    \epsilon_n
    \end{bmatrix},
 \end{eqnarray}   
$
    \X_{i} = \left(X_{i1},X_{i2},\ldots,X_{iL-1}\right)$ e $ 
    \bbeta = \left(\beta_{1},\beta_{2},\ldots,\beta_{L-1}\right)^T$. Suppose that $\bepsilon \sim N(\mathbf{0},\mbox{I}\sigma^{2})$ are random errors. The parameter $\beta_0$ represents the effect of the reference class $L$ and $\beta_l$ represents the increment of class $l$ with respect to the reference class. The interpretation of the model does not change if the reference class changes. If $K= 1$ and $L= 2$ we obtain the model studied by [Buonaccorsi et al.,
2005].

Due to $\X$ being unobservable, an estimate of the parameters is derived from the model with the covariates with classification error, therefore
\begin{eqnarray}
Y_i &=& \gamma_{0} + \sum_{l=1}^{L-1} \gamma_{l} W_{il} + \epsilon_{Wi}, \label{eq2: modelow}
\end{eqnarray} whose matrix form is given by
\begin{eqnarray}
\Y&=& \mathds{1} \gamma_{0} + \W \bgamma + \bepsilon_{W}, 
\end{eqnarray}
where
\begin{eqnarray}
\Y&=&\begin{bmatrix}
    Y_1\\
    \vdots\\
    Y_n
\end{bmatrix},\nonumber
    \W = \begin{bmatrix}
    \W_1\\
    \vdots\\
    \W_n
\end{bmatrix},\bepsilon_W = \begin{bmatrix}
    \epsilon_{W1}\\
    \vdots\\
    \epsilon_{Wn}
    \end{bmatrix},
 \end{eqnarray} 
where $\W_i = \left(W_{i1},W_{i2},\ldots,W_{iL-1}\right)$ e $ 
\bgamma = \left(\gamma_{1},\gamma_{2},\ldots,\gamma_{L-1}\right)$. Suppose that $\bepsilon_W \sim N(\mathbf{0},\mathds{I}\sigma^{2}_W)$ are random errors with variance $\sigma^2_W$. Since the model (\ref{eq2: modelow}) is linear in $\bgamma$ e $\gamma_{0}$, we wil use the least square estimator $\widehat{\bgamma} = \left(\W^T\W\right)^{-1}\W^T\Y$ to estimate $\bgamma$. It is known that $\widehat{\bgamma}$ is an unbiased estimator of $\gamma$; however, the interest lies in obtaining an estimator of $\bbeta$ that is asymptotically unbiased. In order to construct this estimator $\widehat{\bgamma}$, we will use the variance and covariance matrices defined below.

Equation \ref{eq1:modelox} defines that, for all $i$,

\begin{eqnarray*} \label{eq:cov.x.y}
\mbox{Cov}(X_{il},Y_i)&=&\mbox{Cov}(X_{il}, \beta_0  +\sum_{l'=1}^{L-1} \beta_{l'} X_{il'})=\sum_{l'=1}^{L-1}\beta_{l'}\mbox{Cov}(X_{il},X_{il'}) \nonumber \\
 &=& \mathbf{C}_{l} \mathbf{\bbeta}, \mbox{ for } l=1,\ldots,L-1,
\label{eq:CovX}\end{eqnarray*}

\noindent where $\bbeta = \left(\beta_{1},\beta_{2},\ldots,\beta_{L-1}\right)^T$ and $\mathbf{C}_l = \left[ \mbox{Cov}(X_{il}, X_{i1}), \ldots, \mbox{Cov}(X_{il}, X_{iL-1})\right]$ is a vector with dimension $1\times(L-1)$. Defining
\begin{eqnarray*}
\mathbf{\Sigma}_X = \begin{bmatrix}
    \mathbf{C}_1\\
    \vdots\\
    \mathbf{C}_{L-1}\\
\end{bmatrix}
\label{eq:sigmaX}
\end{eqnarray*}
as a covariance matrix with $(L-1)\times(L-1)$ as dimensions, and
\begin{eqnarray*}
\mathbf{\Sigma}_{XY} = \begin{bmatrix}
    \mbox{Cov}(Y_i,X_{i1})\\
    \vdots\\
    \mbox{Cov}(Y_i,X_{iL-1})\\
\end{bmatrix}
\label{eq:sigmaXY}
\end{eqnarray*}

a vector with dimension $(L-1)\times 1$. Note that
\begin{eqnarray}
    \mathbf{\Sigma}_{XY} = \mathbf{\Sigma}_X\bbeta
\label{eq:MSigmaX}.
\end{eqnarray}
Similarly, by \eqref{eq2: modelow}, defining $\mathbf{D}_l = \left[ \mbox{Cov}(W_{il}, W_{i1}), \ldots, \mbox{Cov}(W_{il}, W_{iL-1})\right]$,

\begin{eqnarray*}
\mathbf{\Sigma}_W = \begin{bmatrix}
    \mathbf{D}_1\\
    \vdots\\
    \mathbf{D}_{L-1}\\
\end{bmatrix},
\label{eq:sigmaX}
\end{eqnarray*}
and
\begin{eqnarray*}
\mathbf{\Sigma}_{WY} = \begin{bmatrix}
    \mbox{Cov}(Y_i,W_{i1})\\
    \vdots\\
    \mbox{Cov}(Y_i,W_{iL-1})\\
\end{bmatrix},
\label{eq:sigmaXY}
\end{eqnarray*}
we obtain
\begin{eqnarray}
    \mathbf{\Sigma}_{WY} = \mathbf{\Sigma}_W\bgamma
\label{eq:MSigmaW}.
\end{eqnarray}

Finally, considering $Y_i$ from \eqref{eq1:modelox}, we obtain

\begin{eqnarray*}
\mbox{Cov}(W_{il},Y_{i}) &=& \mbox{Cov}(W_{il},\beta_{0}  +\sum_{l'=1}^{L-1} \beta_{l'} X_{il'})=\sum_{l'=1}^{L-1}\beta_{l'}\mbox{Cov}(W_{il},X_{il'}) \nonumber \\
&=&\left[\begin{array}{ccc}
    \mbox{Cov}(W_{il},X_{i1}) & \ldots & \mbox{Cov}(W_{il},X_{iL-1}) \\
\end{array}\right] \left[\begin{array}{c}
\beta_{1}\\
\vdots\\
\beta_{L-1}
\end{array}\right].
\end{eqnarray*}
and therefore
\begin{eqnarray}
\left[\begin{array}{c}
\mbox{Cov}(W_{i1},Y_i)\\
\vdots\\
\mbox{Cov}(W_{il},Y_i)\\
\vdots\\
\mbox{Cov}(W_{iL-1},Y_i)
\end{array}\right] &=& \left[\begin{array}{ccc}
  \mbox{Cov}(W_{i1},X_{i1}) & \ldots & \mbox{Cov}(W_{iL-1},X_{i1}) \\
   \vdots  & \ddots & \vdots \\
  \mbox{Cov}(W_{i1},X_{iL-1}) & \ldots & \mbox{Cov}(W_{iL-1},X_{iL-1})   
\end{array}\right] \left[\begin{array}{c}
\beta_1\\
\vdots\\
\beta_{L-1}
\end{array}\right]. \nonumber
\end{eqnarray}
Defining

\begin{eqnarray}
\mathbf{\Sigma}_{WX} &=& \left[\begin{array}{ccc}
  \mbox{Cov}(W_{i1},X_{i1}) & \ldots & \mbox{Cov}(W_{iL-1},X_{i1}) \\
   \vdots  & \ddots& \vdots \\
  \mbox{Cov}(W_{i1},X_{iL-1}) & \ldots & \mbox{Cov}(W_{iL-1},X_{iL-1})   
\end{array}\right]\nonumber
\end{eqnarray} it follows that
\begin{eqnarray}
     \mathbf{\Sigma}_{WY} = \mathbf{\Sigma}_{WX}\bbeta 
\label{eq3:sigmawx}.
\end{eqnarray}

Considering the linear model in \eqref{eq2: modelow}, and defining  $\widehat{\bgamma}=(\W^T\W)^{-1}\W^T\Y$ as the least square estimator of $\bgamma$, it follows that
\begin{eqnarray}
\widehat{\bgamma} \overset{P}{\to} \bgamma
\label{eq:aprox}
\end{eqnarray}
that is, $\widehat{\bgamma}$ converges in probability to $\bgamma$.
Using the equation \eqref{eq:MSigmaW}, 
\begin{eqnarray}
\bgamma=\mathbf{\Sigma}^{-1}_W\mathbf{\Sigma}_{WY}.
\label{eq:gamma}
\end{eqnarray}
and substituting \eqref{eq3:sigmawx} in \eqref{eq:gamma}, we obtain $$\bgamma=\mathbf{\Sigma}^{-1}_W\mathbf{\Sigma}_{WX}\bbeta.$$

Thus, the corrected estimator $\widehat{\bbeta}_C$ of $\bbeta$ is given by the correction of $\widehat{\bgamma}$ through the transformation
\begin{eqnarray}
\widehat{\bbeta}_C&=&\left(\mathbf{\Sigma}^{-1}_W\mathbf{\Sigma}_{WX}\right)^{-1}\widehat{\bgamma}.
\label{eq:betachap}
\end{eqnarray}
\subsubsection{Calculation of the variance and covariance matrices}

In order to obtain the correction defined in \eqref{eq:betachap} we need to determine the analytical expressions of the matrices $\bSigma_W$ and $\bSigma_{WX}$. 
Reminder that $\W_{i}= (W_{i1},\ldots,W_{i \, L-1})$, it follows that
\begin{eqnarray}
\mbox{Var}(W_{il}) &=& E[W_{il}^2] - E^2[W_{il}]\nonumber\\
&=& E_X[E_{W|X}[(W_{il}|X_i]]-\left(E_X[E_{W|X}[(W_{il}|X_i]]\right)^2 \nonumber \\
&=& \displaystyle\sum^{L}_{x=1}P(X_i=x)E[W_{il}|X_i=x] - \left(\displaystyle\sum^{L}_{x=1}P(X_i=x)E[W_{il}|X_i=x]\right)^2. \nonumber
\end{eqnarray}
Considering (\ref{eq:theta}), it follows that $$\theta_{l|x}=P(W_i=l|X_i=x)=P(W_{il}=1|X_i=x)=E[W_{il}|X_i=x],$$ in such a way that
\begin{eqnarray}
\mbox{Var}(W_{il})&=& \displaystyle\sum^{L}_{x=1}P(X_i=x)\theta_{l|x} -  \left( \displaystyle\sum^{L}_{x=1}P(X_i=x)\theta_{l|x}\right)^2.
\label{eq:VarW}
\end{eqnarray}
Moreover, $l \ne m$,
\begin{eqnarray}
\mbox{Cov}(W_{il},W_{im}) &=& E[W_{il}W_{im}] - E[W_{il}]E[W_{im}]\nonumber\\
&=& E[W_{il}W_{im}] - \displaystyle\sum^{L}_{x=1}P(X_i=x)\theta_{l|x} \displaystyle\sum^{L}_{x=1}P(X_i=x)\theta_{m|x} \nonumber \\
&=& - \displaystyle\sum^{L}_{x=1}P(X_i=x)\theta_{l|x} \displaystyle\sum^{L}_{x=1}P(X_i=x)\theta_{m|x} ,
\label{eq:CovW}
\end{eqnarray}
where $E[W_{il}W_{im}]=0$,since $W_{il}W_{im}=0$ when $l \ne m$. Reminder that the equations \ref{eq:VarW} and \ref{eq:CovW}, are utilized to construe the matrix $\Sigma_W$.
Note that
\begin{eqnarray}\nonumber
\mbox{Cov}(W_{il'},X_{il}) &=& E[W_{il'}X_{il}] - E[W_{il'}]E[X_{il}] \\
 &=& E[W_{il'}X_{il}] - \left(\sum_{x=1}^LP(X_i=x)\theta_{l'|x}\right) 
 \label{eq:CovIncWX}\end{eqnarray}
\begin{eqnarray}\nonumber
 E_X[E_{W|X}[W_{il'},X_{il}|X_i]] &=& \sum_{x=1}^LP(X_i=x) E[W_{il'},X_{il}|X_i]\\ \nonumber
 &=& \sum_{x=1}^LP(X_i=x) \mbox{I}_{(x=l)} E[W_{il'}|X_i]\\
 &=& P(X_i=l')\theta_{l'|l}.
\label{eq:EWX}\end{eqnarray}

By substituting equation (\ref{eq:EWX}) into equation (\ref{eq:CovIncWX}), we obtain
\begin{eqnarray}\nonumber
\mbox{Cov}(W_{il'},X_{il}) &=& \left(\theta_{l'|l} -\sum_{x=1}^LP(X_i=x)\theta_{l'|x}\right)P(X=l),
\label{eq:CovWX}\end{eqnarray}
which defines an element of the $\Sigma_{WX}$ matrix.

\subsection{The linear regression model with more than one categorical variable having two or more categories with classification error}
\label{mult}

Let \( y_i \), \( i = 1, \ldots, n \), be an observation of the random response variable \( Y_i \), and let \( X_{ik} \), \( k = 1, \ldots, K \), be categorical covariates where the \(k\)-th covariate has \( L_k \) categories. Using the same criterion as in Section \ref{sing}, define
\[
X_{ikl} = 1 \quad \text{if and only if} \quad X_{ik} = l.
\]
Thus, the model has the matrix form
\begin{eqnarray}
    \Y=\mathds{1} \beta_0 + \X \bbeta + \bepsilon \nonumber
\label{eq3:modeloxK}
\end{eqnarray}
where the design matrix is given by $\X = \left( \X_{1}, \ldots,\X_n \right)^T$ with \\ $\X_i = \left(X_{i11},\ldots,X_{i1L_1-1}, \ldots, X_{iK1},\ldots, X_{iKL_K-1}\right)$, $\beta_0$ is the intercept and\\ $\bbeta = \left(\beta_{11},\beta_{12},\ldots,\beta_{K1},\ldots,\beta_{KL-1}\right)$ is the vector parameters associated with $\X_{i}$ with dimension $\sum_{k=1}^{K}(L_k-1)$. Moreover, it follows that $\bepsilon = \left(\epsilon_{1},\epsilon_{2},\ldots,\epsilon_{n}\right)$ is a vector of random errors, such that $\bepsilon \sim N(\mathbf{0},\mathds{I}\sigma^{2})$, where $\mathds{I}$ represents the identity matrix of order $n \times n$. Considering now that the covariates are measured with error, and $\X$ is unobservable, the linear model that can be fitted is given by
\begin{eqnarray}
    \Y&=&\mathds{1} \gamma_{0}   + \W\bgamma + \bepsilon_W,
\label{eq3:modelowK}
\end{eqnarray}
where
$$\bgamma = \left(\gamma_{11},\gamma_{12},\ldots,\gamma_{1L_1-1}, \gamma_{21},\ldots,\gamma_{KL_K-1}\right) $$ is a vector with dimension $\sum_{k=1}^{K}(L_k-1)$  e
$\bepsilon_W = \left(\epsilon_{W1},\epsilon_{W2},\ldots,\epsilon_{Wn}\right),$ is a vector of random errors with length $n$, such that $\bepsilon_W \sim N(\mathbf{0},I\sigma^{2}_W)$. Moreover, the design matrix $\W = (\mathbf{W_1}, \ldots, \mathbf{W_n})^T$ has components
$\W_i = \left(W_{i11},W_{i12},\ldots,W_{i1L_1-1}, W_{i21},\ldots, W_{iKL_K-1}\right)$.

In order to construct an asymptotically unbiased estimator $\widehat{\bbeta}$, we need to calculate the following covariances, such that, for $Y_i$ defined in\eqref{eq3:modelowK}
\begin{eqnarray}
    \mbox{Cov}(W_{ikl},Y_i) &=& \mbox{Cov}(\gamma_{0}   + \sum_{k'=1}^K\sum_{l'=1}^{L_{k'}-1}W_{ik'l'}\gamma_{k'l'} +  \epsilon_{Wi}, W_{ikl})\nonumber\\
        & = & \sum_{k'=1}^K\sum_{l'=1}^{L_{k'}-1}\mbox{Cov}(W_{ik'l'},W_{ikl})\gamma_{k'l'}\nonumber\\
        &=& \sum_{l'=1}^{L_{k}-1}\mbox{Cov}(W_{ikl'},W_{ikl})\gamma_{kl'}
\end{eqnarray}
because $\mbox{Cov}(W_{ikl},W_{ik'l'}) =  0 \text{ if $k \ne k'$}.$ Therefore
\begin{eqnarray*}
    \mbox{Cov}(W_{ikl},Y_i) &=& \Big[ \mbox{Cov}(W_{ikl},W_{ik1}) \quad \ldots \quad \mbox{Cov}(W_{ikl},W_{ikL_{k}-1})
     \Big] \bgamma_{k},
\end{eqnarray*}

where
\begin{eqnarray*}
\bgamma_{k} &=& \begin{bmatrix}
\gamma_{k1}\\
\gamma_{k2}\\
\vdots\\
\gamma_{kL-1}
\end{bmatrix}
\end{eqnarray*}
and
\begin{eqnarray}
\mathbf{\Sigma}_{W_{k}} &=&  \begin{bmatrix}
\mbox{Cov}(W_{ik1},W_{ik1}) & \ldots & \mbox{Cov}(W_{ik1},W_{ikL-1})\\
\mbox{Cov}(W_{ik2},W_{ik1}) &  \ldots & \mbox{Cov}(W_{ik2},W_{ikL-1})\\
\vdots & \ddots & \vdots\\
\mbox{Cov}(W_{ikL-1},W_{ik1}) & \ldots & \mbox{Cov}(W_{ikL-1},W_{ikL-1})
\end{bmatrix} \label{eq:sigwk}.
\end{eqnarray} 

Let
\begin{eqnarray*}
    \mathbf{\Sigma}_{WY} &=& \begin{bmatrix}
        \mbox{Cov}(Y_i, W_{i11})\\
        \vdots\\
        \mbox{Cov}(Y_i, W_{i1L_1-1})\\
        \vdots\\
        \mbox{Cov}(Y_i, W_{iKL_K1})\\
        \vdots\\
        \mbox{Cov}(Y_i, W_{iKL_k-1})
    \end{bmatrix},
\end{eqnarray*}
    
\begin{eqnarray*}
    \mathbf{\Sigma}_{W} &=& \begin{bmatrix}
     \mathbf{\Sigma}_{W_1}    & \mathbf{\undertilde{0}} & \ldots & \mathbf{\undertilde{0}}  \\
     \mathbf{\undertilde{0}}   &  \mathbf{\Sigma}_{W_2} & \ldots & \mathbf{\undertilde{0}}\\
     \vdots & & \ddots & \vdots\\
     \mathbf{\undertilde{0}}   &  \mathbf{\undertilde{0}}   &  \ldots & \mathbf{\Sigma}_{W_K}
    \end{bmatrix},
\end{eqnarray*}

\begin{eqnarray*}
\bbeta &=& \begin{bmatrix}
 \bbeta_{11} \\
 \bbeta_{12} \\
 \vdots\\
 \bbeta_{K1} \\
 \vdots\\
 \bbeta_{KL_K-1}
\end{bmatrix} \mbox{ e}
\end{eqnarray*}

\begin{eqnarray*}
\bgamma &=& \begin{bmatrix}
 \bgamma_{1} \\
 \bgamma_{2} \\
 \vdots\\
 \bgamma_{k} \\
\end{bmatrix}.
\end{eqnarray*}
Similarly as Equation (\ref{eq:MSigmaX}) , it follows that \begin{eqnarray}
  \mathbf{\bSigma}_{WY} = \mathbf{\bSigma}_{W}\bgamma.
   \label{eq4:sigmawy}  
\end{eqnarray} 

Defining $\bSigma_{W_kX}$ as

\begin{eqnarray*}
    \mathbf{\bSigma}_{W_KX} &=& \begin{bmatrix}
        \mbox{Cov}(X_{ik1},W_{ik1}) & \mbox{Cov}(X_{ik2},W_{ik1})  & \ldots & \mbox{Cov}(X_{ikL_k-1},W_{ik1}) \\
        \mbox{Cov}(X_{ik1},W_{ik2}) & \mbox{Cov}(X_{ik2},W_{ik2})  & \ldots & \mbox{Cov}(X_{ikL_k-1},W_{ik2}) \\
        \vdots & & \ddots & \vdots \\
        \mbox{Cov}(X_{ik1},W_{ikL_k-1}) & \mbox{Cov}(X_{ik2},W_{ikL_k-1})  & \ldots & \mbox{Cov}(X_{ikL_k-1},W_{ikL_k-1})
    \end{bmatrix},
\end{eqnarray*}
we obtain,
\begin{eqnarray*}
    \mathbf{\bSigma}_{WX} &=&  \begin{bmatrix}
     \mathbf{\bSigma}_{W_{1X}}    & \mathbf{\undertilde{0}} & \ldots & \mathbf{\undertilde{0}}  \\
     \mathbf{\undertilde{0}}   &  \mathbf{\bSigma}_{W_{2X}} & \ldots & \mathbf{\undertilde{0}}\\
     \vdots & & \ddots & \vdots\\
     \mathbf{\undertilde{0}}   &  \mathbf{\undertilde{0}}   &  \ldots & \mathbf{\bSigma}_{W_{KX}} 
    \end{bmatrix} 
\end{eqnarray*}
and, similarly as \eqref{eq3:sigmawx},
\begin{eqnarray}
    \mathbf{\bSigma}_{WY} &=& \mathbf{\bSigma}_{WX}\bbeta.
    \label{eq5:sigmawyx}
\end{eqnarray}

Substituting (\ref{eq5:sigmawyx}) in (\ref{eq4:sigmawy}) we obtain

\begin{eqnarray}
\label{eq:sigwy}
    \mathbf{\bSigma}_{WY} &= \mathbf{\bSigma}_{WX}\bbeta =& \mathbf{\bSigma}_{W}\bgamma,
\end{eqnarray}
such that $\widehat{\bbeta}_C = (\bSigma_W^{-1}\bSigma_{WX})^{-1}\widehat{\bgamma}$. 

The components of $\bSigma_{WX}$ and of $\bSigma_{W}$ are calculated the same way as (\ref{eq:VarW}), (\ref{eq:CovW}) e (\ref{eq:CovWX}).

\subsection{$\beta_0$ correction}\label{sec:intercept}

Note that the regression described in \eqref{eq3:modelowK} has \(\sum_{k=1}^K k(L_k - 1) + 1\) parameters to be estimated. However, due to the singularity of the covariance matrix of multinomial variables defined in \eqref{eq:sigwk}, the method described in Section \ref{mult} corrects only the vector of estimators \(\widehat{\boldsymbol{\gamma}}\), giving rise to the vector \(\widehat{\boldsymbol{\beta}}_C\) defined in \eqref{eq:betachap} and \eqref{eq:sigwy}, with dimension \(\left(\sum_{k=1}^K k(L_k - 1)\right) \times 1\). Thus, the intercept \(\widehat{\gamma}_0\) still presents a bias.

The objective is to correct the bias of the estimator \(\widehat{\gamma}_0\), obtained by the least squares method through the regression of \(\mathbf{Y}\) on \(\mathbf{W}\). Given the random variable \(\mathbf{X}\), and
\[
\mathbf{Y} = \beta_0 + \mathbf{X}\boldsymbol{\beta} + \epsilon,
\]
such that
\[
E[\mathbf{Y} \mid \mathbf{X}] = \beta_0 + \mathbf{X}\boldsymbol{\beta}
\]
and
\[
E[\mathbf{Y} \mid \mathbf{W}] = \beta_0 + \boldsymbol{\beta} E[\mathbf{X} \mid \mathbf{W}] = \beta_0 + \boldsymbol{\beta} \boldsymbol{\pi},
\]
where
\[
\boldsymbol{\pi} = \begin{bmatrix}
\pi_{1|W_1} & \ldots & \pi_{L_k|W_1} \\
\pi_{1|W_2} & \ldots & \pi_{L_k|W_2} \\
\vdots & \ddots & \vdots \\
\pi_{1|W_n} & \ldots & \pi_{L_k|W_n}
\end{bmatrix},
\]
we propose the corrected estimator \(\widehat{\beta}_{0C}\) for \(\beta_0\), given by
\begin{equation}
\label{eq:beta0}
\widehat{\beta}_{0C} = \frac{\displaystyle\sum_{i=1}^n \left(Y_i - \boldsymbol{\pi}_{(i)} \widehat{\boldsymbol{\beta}}_C \right)}{n},
\end{equation}
where \(\boldsymbol{\pi}_{(i)}\) is the \(i\)-th row of the matrix \(\boldsymbol{\pi}\), and \(Y_i\) is the \(i\)-th observation of the vector \(\mathbf{Y}\).

\section{Calculation of estimator biases}\label{sec:viés}
Let
\[
\boldsymbol{\beta}^* = (\beta_0, \boldsymbol{\beta})^T,
\]
with dimension \(M = \sum_{k=1}^K(L_k - 1) + 1\), be the parameter vector of the regression of \(\mathbf{Y}\) on \(\mathbf{X}^* = (\mathbf{1}, \mathbf{X})\) as given by \eqref{eq1:modelox}, and let \(\widehat{\boldsymbol{\beta}}_C^* = (\widehat{\gamma}_0, \widehat{\boldsymbol{\beta}}_C)^T\), where \(\widehat{\boldsymbol{\beta}}_C\) is the corrected vector as described in Section~\ref{mult}. Furthermore, let \(\boldsymbol{\gamma}^* = (\gamma_0, \boldsymbol{\gamma})\) be the parameter vector of the regression of \(\mathbf{Y}\) on \(\mathbf{W}^* = (\mathbf{1}, \mathbf{W})\) as given by \eqref{eq3:modelowK}. Consider

\[
\widehat{\boldsymbol{\gamma}}^* = (\widehat{\gamma}_0, \widehat{\boldsymbol{\gamma}}) = (\mathbf{W}^{*T} \mathbf{W}^*)^{-1} \mathbf{W}^{*T} \mathbf{Y}
\]

as the vector of estimators of \(\boldsymbol{\gamma}^*\).

To find the conditional bias of \(\widehat{\boldsymbol{\beta}}_C^*\) given \(\mathbf{W}\), we compute:

\[
\mathbb{E}_{\mathbf{Y}|\mathbf{W}}\left[\widehat{\boldsymbol{\beta}}_C^*|\mathbf{W}\right] = \mathbb{E}_{\mathbf{Y}|\mathbf{W}}\left[\mathbf{Z} \widehat{\boldsymbol{\gamma}}^*|\mathbf{W}\right],
\]

where

\[
\mathbf{Z} = \begin{bmatrix}
1 & \mathbf{0}^T \\
\mathbf{0} & \left(\boldsymbol{\Sigma}_W^{-1} \boldsymbol{\Sigma}_{WX}\right)^{-1}
\end{bmatrix}
\]

is a matrix of dimension \(M \times M\). Thus,

\begin{align*}
\mathbb{E}_{\mathbf{Y}|\mathbf{W}}[\mathbf{Z} \widehat{\boldsymbol{\gamma}}^* | \mathbf{W}] 
&= \mathbf{Z} \mathbb{E}_{\mathbf{Y}|\mathbf{W}}[\widehat{\boldsymbol{\gamma}}^* | \mathbf{W}] \\
&= \mathbf{Z} (\mathbf{W}^{*T} \mathbf{W}^*)^{-1} \mathbf{W}^{*T} \mathbb{E}_{\mathbf{Y}|\mathbf{W}}[\mathbf{Y} | \mathbf{W}] \\
&= \mathbf{Z} (\mathbf{W}^{*T} \mathbf{W}^*)^{-1} \mathbf{W}^{*T} \mathbb{E}_{\mathbf{X}|\mathbf{W}}[\mathbf{X}^* \boldsymbol{\beta}^* | \mathbf{W}] \\
&= \mathbf{Z} (\mathbf{W}^{*T} \mathbf{W}^*)^{-1} \mathbf{W}^{*T} \mathbb{E}_{\mathbf{X}|\mathbf{W}}[\mathbf{X}^* | \mathbf{W}] \boldsymbol{\beta}^*.
\end{align*}

Let

\[
\mathbb{E}_{\mathbf{X}|\mathbf{W}}[\mathbf{X}^* | \mathbf{W}] = 
\begin{bmatrix}
1 & \pi_{1|w_1} & \cdots & \pi_{M|w_1} \\
1 & \pi_{1|w_2} & \cdots & \pi_{M|w_2} \\
\vdots & \vdots & \ddots & \vdots \\
1 & \pi_{1|w_n} & \cdots & \pi_{M|w_n}
\end{bmatrix} = \boldsymbol{\pi}^*.
\]

Therefore,

\begin{equation}
\mathbb{E}_{\mathbf{Y}|\mathbf{W}}\left[\widehat{\boldsymbol{\beta}}_C^*|\mathbf{W}\right] = \mathbf{Z}(\mathbf{W}^{*T} \mathbf{W}^*)^{-1} \mathbf{W}^{*T} \boldsymbol{\pi}^* \boldsymbol{\beta}^*,
\label{eq:viésbeta}
\end{equation}

and the bias \(\mathbf{B}\) of \(\widehat{\boldsymbol{\beta}}_C^*\) is given by:

\begin{align*}
\mathbf{B} &= \mathbb{E}_{\mathbf{Y}|\mathbf{W}}\left[\widehat{\boldsymbol{\beta}}_C^*|\mathbf{W}\right] - \boldsymbol{\beta}^* \\
&= \left( \mathbf{Z}(\mathbf{W}^{*T} \mathbf{W}^*)^{-1} \mathbf{W}^{*T} \boldsymbol{\pi}^* - \mathds{I} \right) \boldsymbol{\beta}^*,
\end{align*}

where \(\mathds{I}\) is the identity matrix of dimension \(M \times M\). Note that when the conditional probabilities satisfy

\[
\pi_{j|w_k} \to
\begin{cases}
1, & \text{if } j = w_k \\
0, & \text{otherwise}
\end{cases}
\]

we have \(\boldsymbol{\pi}^* \to \mathbf{W}^*\) and \(\mathbf{Z} \to \mathds{I}\), and thus \(\mathbf{B} \to \mathbf{0}\).

The estimator \(\widehat{\beta}_{0C}\) of the intercept \(\beta_0\) in the regression of \(\mathbf{Y}\) on \(\mathbf{X}\), is determined independently from the other estimators via \eqref{eq:beta0}. Consequently, we compute its bias independently. Let

\[
\mathbb{E}_{\mathbf{Y}|\mathbf{W}}\left[\widehat{\beta}_{0C}|\mathbf{W}\right] = \beta_0 + B_0,
\]

where

\begin{align}
\mathbb{E}_{\mathbf{Y}|\mathbf{W}}\left[\widehat{\beta}_{0C}|\mathbf{W}\right] 
&= \mathbb{E}_{\mathbf{Y}|\mathbf{W}}\left[\frac{1}{n} \sum_{i=1}^n \left(Y_i - \boldsymbol{\pi}_{(i)} \widehat{\boldsymbol{\beta}}_C \right) \Big| \mathbf{W} \right] \notag \\
&= \frac{1}{n} \sum_{i=1}^n \left( \mathbb{E}_{\mathbf{Y}|\mathbf{W}}[Y_i | \mathbf{W}] - \boldsymbol{\pi}_{(i)} \mathbb{E}_{\mathbf{Y}|\mathbf{W}}\left[\widehat{\boldsymbol{\beta}}_C | \mathbf{W} \right] \right) \notag \\
&= \frac{1}{n} \sum_{i=1}^n \left( \mathbb{E}_{\mathbf{X}|\mathbf{W}}[(1, \mathbf{X}_i) | \mathbf{W}] \boldsymbol{\beta}^* - \boldsymbol{\pi}_{(i)} \mathbb{E}_{\mathbf{Y}|\mathbf{W}}[\widehat{\boldsymbol{\beta}}_C | \mathbf{W}] \right) \notag \\
&= \frac{1}{n} \sum_{i=1}^n \left( \boldsymbol{\pi}_{(i)}^* \boldsymbol{\beta}^* - \boldsymbol{\pi}_{(i)} \mathbb{E}_{\mathbf{Y}|\mathbf{W}}[\widehat{\boldsymbol{\beta}}_C | \mathbf{W}] \right).
\label{eq:viésB0}
\end{align}

Here, \(\boldsymbol{\pi}_{(i)}\) and \(\boldsymbol{\pi}_{(i)}^*\) are the \(i\)-th rows of matrices \(\boldsymbol{\pi}\) and \(\boldsymbol{\pi}^*\), respectively. Note that \\ \(\mathbb{E}_{\mathbf{Y}|\mathbf{W}}[\widehat{\boldsymbol{\beta}}_C | \mathbf{W}]\) is the vector \(\mathbb{E}_{\mathbf{Y}|\mathbf{W}}[\widehat{\boldsymbol{\beta}}_C^* | \mathbf{W}]\) without the first element. Then,

\begin{align*}
\mathbb{E}_{\mathbf{Y}|\mathbf{W}}\left[\widehat{\beta}_{0C}|\mathbf{W}\right] 
&= \frac{1}{n} \sum_{i=1}^n \left( \beta_0 + \boldsymbol{\pi}_{(i)} \boldsymbol{\beta} - \mathbb{E}_{\mathbf{Y}|\mathbf{W}}\left[\widehat{\boldsymbol{\beta}}_C | \mathbf{W} \right] \right) \\
&= \beta_0 + \frac{1}{n} \sum_{i=1}^n \left( \boldsymbol{\pi}_{(i)} \boldsymbol{\beta} - \mathbb{E}_{\mathbf{Y}|\mathbf{W}}\left[\widehat{\boldsymbol{\beta}}_C | \mathbf{W} \right] \right) \\
&= \beta_0 + B_0.
\end{align*}

Hence,

\[
B_0 = \frac{1}{n} \sum_{i=1}^n \left( \boldsymbol{\pi}_{(i)} \boldsymbol{\beta} - \mathbb{E}_{\mathbf{Y}|\mathbf{W}}\left[\widehat{\boldsymbol{\beta}}_C | \mathbf{W} \right] \right).
\]

\section{Calculation of variance of estimators}\label{seq:var}

Let $\bgamma^* = (\gamma_0, \bgamma)$ be the parameter vector of the regression of $\Y$ on $\W^*=(\one, \W)$ given by Equation \eqref{eq2: modelow}. And let
\[
\widehat{\bgamma}^*=(\widehat{\gamma}_0, \widehat{\bgamma})= \left( \W^{*T}\W^* \right)^{-1} \W^{*T} \Y.
\]
The variance of $\widehat{\bgamma}^*$ is calculated below. Consider
\begin{align}
    \Var(\widehat{\bgamma}^*) &= \Var\left((\W^{*T}\W^*)^{-1}\W^{*T}\Y\right)\nonumber\\
    &= (\W^{*T}\W^*)^{-1}\W^{*T} \Var\left(\Y\right)\W^{*}(\W^{*T}\W^*)^{-1}\nonumber\\
    &= (\W^{*T}\W^*)^{-1}\sigma_W^2.
\end{align}

Let $\widehat{\bbeta}_C^*=(\widehat{\gamma}_{0}, \widehat{\bbeta}_C)^T$ and define
\[
\mathbf{Z}=\begin{bmatrix}
1 & \mathbf{0}^T \\
\mathbf{0} & \left(\Sigma_W^{-1}\Sigma_{WX}\right)^{-1}
\end{bmatrix},
\]
such that
\begin{align}
    \Var\left(\widehat{\bbeta}^*_C\right) &= \Var\left(\Z\widehat{\bgamma}^*\right)\nonumber\\
    &= \Var\left(\Z\left(\W^{*T}\W^*\right)^{-1}\W^{*T}\Y\right) \nonumber\\
    &= \Z\left(\W^{*T}\W^*\right)^{-1}\W^{*T}\Var\left(\Y\right)\W^{*}\left(\W^{*T}\W^*\right)^{-1}\Z^T\nonumber \\
    &= \sigma_W^2 \Z\left(\W^{*T}\W^*\right)^{-1}\Z^T.
\end{align}

Therefore, by the construction of $\Z$, $\Var\left(\widehat{\bbeta}_C\right)$ is equal to $\Var\left(\widehat{\bbeta}^*_C\right)$ without the first element of the vector.

Using Equation \eqref{eq:beta0}, we obtain
\begin{align}
    \Var\left(\widehat{\beta}_{0C}\right) &= \Var\left(\frac{\displaystyle\sum_{i=1}^n\left(Y_i-\bpi_{(i)}\widehat{\bbeta}_C\right)}{n}\right)\nonumber\\
    &= \Var\left(\frac{\displaystyle\sum_{i=1}^n\left(Y_i-\bpi_{(i)}\left(\Sigma_W^{-1}\Sigma_{WX}\right)^{-1}\widehat{\bgamma}\right)}{n}\right)\nonumber\\
    &= \frac{1}{n^2}\sum_{i=1}^n\left[\Var(Y_i)+\Var\left(\bpi_{(i)}\left(\Sigma_W^{-1}\Sigma_{WX}\right)^{-1}\widehat{\bgamma}\right) - 2\Cov\left(Y_i,\bpi_{(i)}\left(\Sigma_W^{-1}\Sigma_{WX}\right)^{-1}\widehat{\bgamma}\right)\right].
\end{align}

To compute $\Cov\left(Y_i,\bpi_{(i)}\left(\Sigma_W^{-1}\Sigma_{WX}\right)^{-1}\widehat{\bgamma}\right)$, we define $\widehat{\bgamma}$ in terms of $\Y$. Using \eqref{eq2: modelow}, the least squares solution is
\[
\underset{\gamma_0,\bgamma}{\text{minimize}} \quad Q(\gamma_0,\bgamma)=\sum_{i=1}^n\left(Y_i-\gamma_0-\w_i\bgamma\right)^2,
\]
where $\w_i$ is the $i$-th row of $\W$.

Then,
\[
\frac{\partial Q}{\partial \gamma_0}= -2\sum_{i=1}^n\left(Y_i-\gamma_0-\w_i\bgamma\right).
\]
Setting this derivative to zero, we obtain
\begin{eqnarray}
 \gamma_0=\frac{\sum_{i=1}^n\left(Y_i-\w_i\bgamma\right)}{n}. \label{gamma0}   
\end{eqnarray}

Similarly, using \ref{gamma0},
\begin{align*}
    \frac{\partial Q}{\partial \bgamma} &= -2\sum_{i=1}^n\left(Y_i-\gamma_0-\w_i\bgamma\right)\w_i^T\\
    &= -2\sum_{i=1}^n\left(Y_i-\frac{\sum_{j=1}^n\left(Y_j-\w_j\bgamma\right)}{n}-\w_i\bgamma\right)\w_i^T.
\end{align*}

Setting the derivative to zero:
\begin{align*}
    \sum_iY_i\w_i^T - \sum_i\sum_j\frac{(Y_j-\w_j\widehat{\bgamma})}{n}\w_i^T - \sum_i\w_i\widehat{\bgamma}\w_i^T &= \mathbf{0},\\
    \sum_i(Y_i-\bar{Y})\w_i^T &= \A\widehat{\bgamma},
\end{align*}
where $\A=\sum_{i=1}^n\left(\w_i^T\left(\w_i-\frac{\sum_j\w_j}{n}\right)\right)$.

Thus,
\begin{eqnarray}
 \widehat{\bgamma} = \A^{-1}\left(\sum_{k=1}^n\left(Y_k-\bar{Y}\right)\w_k^T\right). \label{vargammachap}   
\end{eqnarray}

Substituting \ref{vargammachap} into $\Var(\bpi_{(i)}\left(\Sigma_W^{-1}\Sigma_{WX}\right)^{-1}\widehat{\bgamma})$, we get
\begin{align}
\Var\left(\bpi_{(i)}\left(\Sigma_W^{-1}\Sigma_{WX}\right)^{-1}\widehat{\bgamma}\right)=& \nonumber\\  \bpi_{(i)}\left(\Sigma_W^{-1}\Sigma_{WX}\right)^{-1} \A^{-1} \Var\left(\sum_{k=1}^n\left(Y_k-\bar{Y}\right)\w_k^T\right)\A^{-1^T}\left(\left(\Sigma_W^{-1}\Sigma_{WX}\right)^{-1}\right)^T \bpi_{(i)}^T.
\end{align}

Define $V_i = \bpi_{(i)}\left(\Sigma_W^{-1}\Sigma_{WX}\right)^{-1}$, and since
\[
\sum_k\left(Y_k-\bar{Y}\right)\w_k^T = \sum_k\left(\w_k^T-\frac{1}{n}\sum_{l=1}^n\w_l^T\right)Y_k,
\]
we have
\begin{align}
\Var\left(\bpi_{(i)}\left(\Sigma_W^{-1}\Sigma_{WX}\right)^{-1}\widehat{\bgamma}\right) &= \sigma^2 \left[V_i \A^{-1}\left(\sum_k\left(\w_k^T-\frac{1}{n}\sum_l\w_l^T\right)\left(\w_k-\frac{1}{n}\sum_l\w_l\right)\right)\A^{-1^T}V_i^T\right]. \label{vargamma}
\end{align}

Moreover, substituting \ref{vargammachap} into $\Cov\left(Y_i,\bpi_{(i)}\left(\Sigma_W^{-1}\Sigma_{WX}\right)^{-1}\widehat{\bgamma}\right)$:
\begin{align}
\Cov\left(Y_i,\bpi_{(i)}\left(\Sigma_W^{-1}\Sigma_{WX}\right)^{-1}\widehat{\bgamma}\right) &= \Cov\left(Y_i,V_i\A^{-1}\left(\sum_{k=1}^n\left(\Y_k-\bar{Y}\right)\w_k^T\right)\right)\nonumber\\
&= \Cov\left(Y_i,\A^{-1}\left(\sum_{k=1}^n\left(\Y_k-\bar{Y}\right)\w_k^T\right)\right)\A^{-1^T}V_i^T\nonumber\\
&= \sigma^2\left(\w_i-\frac{1}{n}\sum_l\w_l\right)\A^{-1^T}V_i^T. \label{covgamma}
\end{align}

In conclusion, using Equations \ref{vargammachap}, \ref{vargamma}, and \ref{covgamma}, we can compute the variance of the corrected estimator $\widehat{\beta}_{0C}$.

\section{Simulation}
\label{Sim}
We conducted a simulation study to evaluate the correction method for the least squares estimators of the regression model using the observed covariates $\W$, as well as the precision of these estimators. Additionally, the intercept correction method defined in \eqref{eq:beta0} is evaluated by comparing its results with the true value and with the uncorrected case. The evaluation is carried out by comparing the weighted mean squared error defined in \eqref{eq:EQP} across three estimation methods: naive regression without correction for $\bbeta$ and $\beta_0$ (no correction), correction for both $\bbeta$ and $\beta_0$ (full correction), and correction only for $\bbeta$ (partial correction). Furthermore, this study aims to investigate the effects of the number of observations, standard deviation of the response variable, number of categorical variables, magnitude of the components of $\btheta$, and the number of categories on the correction quality in $\widehat{\bbeta}_C$.

We define the number of variables as $K=1,3,10,30,50$, number of categories as $L_k = 2,3,4$ for each $k = 1, \ldots, K$, and the number of observations as $n = \{50, 75, 100, \ldots, 500\}$.  
Moreover, we define $P(W=w|X=x)$, which composes the elements of $\btheta$, under three scenarios (low distortion, medium distortion, and high distortion).  
\newpage
\noindent 1.{ \bf Pouca distorção:}
 
\begin{eqnarray}
\btheta = \begin{cases}
\begin{tabular}{lllllll}
\cline{2-3}
\multicolumn{1}{l|}{}      & \multicolumn{1}{l|}{W=0} & \multicolumn{1}{l|}{W=1} &  &  &  &  \\ \cline{1-3}
\multicolumn{1}{|l|}{X=0} & \multicolumn{1}{l|}{0.9}  & \multicolumn{1}{l|}{0.1}  &  &  &  &  \\ \cline{1-3}
\multicolumn{1}{|l|}{X=1} & \multicolumn{1}{l|}{0.15}  & \multicolumn{1}{l|}{0.85}  &  &  &  &  \\ \cline{1-3}
                           &                           &                           &  &  &  &  \\
                           &                           &                           &  &  &  & 
\end{tabular} \text{ se } L_K=2,\\  
\begin{tabular}{lllllll}
\cline{2-4}
\multicolumn{1}{l|}{}      & \multicolumn{1}{l|}{W=0} & \multicolumn{1}{l|}{W=1} & \multicolumn{1}{l|}{W=2} &  &  &  \\ \cline{1-4}
\multicolumn{1}{|l|}{X=0} & \multicolumn{1}{l|}{0.85} & \multicolumn{1}{l|}{0.1}  & \multicolumn{1}{l|}{0.05} &  &  &  \\ \cline{1-4}
\multicolumn{1}{|l|}{X=1} & \multicolumn{1}{l|}{0.1}  & \multicolumn{1}{l|}{0.8}  & \multicolumn{1}{l|}{0.1}  &  &  &  \\ \cline{1-4}
\multicolumn{1}{|l|}{X=2} & \multicolumn{1}{l|}{0.05} & \multicolumn{1}{l|}{0.1}  & \multicolumn{1}{l|}{0.85} &  &  &  \\ \cline{1-4}
                           &                           &                           &                           &  &  & 
\end{tabular} \text{ se } L_K=3 \mbox{ e}\\

\begin{tabular}{l|l|l|l|l|ll}
\cline{2-5}
                           & W=0  & W=1  & W=2  & W=3  &  &  \\ \cline{1-5}
\multicolumn{1}{|l|}{X=0} & 0.825 & 0.1   & 0.05  & 0.025 &  &  \\ \cline{1-5}
\multicolumn{1}{|l|}{X=1} & 0.075 & 0.8   & 0.075 & 0.05  &  &  \\ \cline{1-5}
\multicolumn{1}{|l|}{X=2} & 0.05  & 0.075 & 0.8   & 0.075 &  &  \\ \cline{1-5}
\multicolumn{1}{|l|}{X=3} & 0.025 & 0.05  & 0.1   & 0.825 &  &  \\ \cline{1-5}
\end{tabular} \text{ se } L_K=4.\\

\end{cases}. 
\end{eqnarray} 
\newpage
\noindent 2.{ \bf Média distorção:}

\begin{eqnarray}
\btheta = \begin{cases}
\begin{tabular}{lllllll}
\cline{2-3}
\multicolumn{1}{l|}{}      & \multicolumn{1}{l|}{W=0} & \multicolumn{1}{l|}{W=1} &  &  &  &  \\ \cline{1-3}
\multicolumn{1}{|l|}{X=0} & \multicolumn{1}{l|}{0.7}  & \multicolumn{1}{l|}{0.3}  &  &  &  &  \\ \cline{1-3}
\multicolumn{1}{|l|}{X=1} & \multicolumn{1}{l|}{0.35}  & \multicolumn{1}{l|}{0.65}  &  &  &  &  \\ \cline{1-3}
                           &                           &                           &  &  &  &  \\
                           &                           &                           &  &  &  & 
\end{tabular} \text{ se } L_K=2,\\  
\begin{tabular}{lllllll}
\cline{2-4}
\multicolumn{1}{l|}{}      & \multicolumn{1}{l|}{W=0} & \multicolumn{1}{l|}{W=1} & \multicolumn{1}{l|}{W=2} &  &  &  \\ \cline{1-4}
\multicolumn{1}{|l|}{X=0} & \multicolumn{1}{l|}{0.7} & \multicolumn{1}{l|}{0.2}  & \multicolumn{1}{l|}{0.1} &  &  &  \\ \cline{1-4}
\multicolumn{1}{|l|}{X=1} & \multicolumn{1}{l|}{0.15}  & \multicolumn{1}{l|}{0.7}  & \multicolumn{1}{l|}{0.15}  &  &  &  \\ \cline{1-4}
\multicolumn{1}{|l|}{X=2} & \multicolumn{1}{l|}{0.1} & \multicolumn{1}{l|}{0.2}  & \multicolumn{1}{l|}{0.7} &  &  &  \\ \cline{1-4}
                           &                           &                           &                           &  &  & 
\end{tabular} \text{ se } L_K=3 \mbox{ e}\\

\begin{tabular}{l|l|l|l|l|ll}
\cline{2-5}
                           & W=0  & W=1  & W=2  & W=3  &  &  \\ \cline{1-5}
\multicolumn{1}{|l|}{X=0} & 0.6 & 0.2   & 0.125  & 0.075 &  &  \\ \cline{1-5}
\multicolumn{1}{|l|}{X=1} & 0.15 & 0.6   & 0.15 & 0.1  &  &  \\ \cline{1-5}
\multicolumn{1}{|l|}{X=2} & 0.1  & 0.15 & 0.6   & 0.15 &  &  \\ \cline{1-5}
\multicolumn{1}{|l|}{X=3} & 0.075 & 0.125  & 0.2   & 0.6 &  &  \\ \cline{1-5}
\end{tabular} \text{ se } L_K=4.\\
 
\end{cases}. 
\end{eqnarray} 
\\
\\
\noindent 3.{ \bf Alta distorção:}

\begin{eqnarray}
\btheta = \begin{cases}
\begin{tabular}{l|l|l|l|l|ll}
\cline{2-5}
                           & W=0  & W=1  & W=2  & W=3  &  &  \\ \cline{1-5}
\multicolumn{1}{|l|}{X=0} & 0.3 & 0.25   & 0.25  & 0.2 &  &  \\ \cline{1-5}
\multicolumn{1}{|l|}{X=1} & 0.25 & 0.3   & 0.25 & 0.2  &  &  \\ \cline{1-5}
\multicolumn{1}{|l|}{X=2} & 0.2  & 0.25 & 0.3   & 0.25 &  &  \\ \cline{1-5}
\multicolumn{1}{|l|}{X=3} & 0.2 & 0.25  & 0.25   & 0.3 &  &  \\ \cline{1-5}
\end{tabular} \text{ sendo } L_K=4.\nonumber\\ 
\end{cases}. 
\end{eqnarray} 
\\
In all scenarios, with all initial parameters defined, we simulate the design matrix $\X$ and then, to simulate observational error, we generate the design matrix $\W$ using $\btheta$, such that
\[
W_i | X_i = x \sim \text{Mult}(1, \btheta_x),
\]
where $\theta_x$ is the $x$-th row of the matrix $\btheta$. Additionally, we transform $\btheta$ into $\bpi$ as described in (\ref{eq:pi_i}). Finally, we define
\[
\beta_l = 0.5 + 0.2l \quad \text{for all } l \in \{0, 1, \ldots, 1 + \sum_{k=1}^K (L_k - 1)\}.
\]

The design matrices $\X$ and $\W$ were initially generated for sample sizes $n=500$ and were then reduced for smaller sample sizes, ensuring nested samples (e.g., the sample of size 400 is a subsample of size 500).

After all preparations, we use $\bbeta$ to simulate the response variable
\[
\Y | \W \sim N(\X \bbeta, \sigma^2 \mathds{I}),
\]
where $\sigma = 0.1, 0.2, 0.5, 1$. For the high distortion scenario, results are only presented for $\sigma = 0.1$ and $1$, since other cases provided no additional relevant information.

We perform least squares regression using design matrix $\W$ in \eqref{eq2: modelow} and \eqref{eq3:modelowK} to obtain estimates $\widehat{\bgamma}$ and $\widehat{\gamma}_0$. Then, the proposed correction method is applied to obtain $\widehat{\bbeta}_C$. Finally, $\widehat{\gamma}_0$ is corrected to obtain $\widehat{\beta}_{C0}$. This process is repeated $M=300$ times, and the average weighted mean squared error is used to compare estimator effectiveness. The weighted mean squared error is given by
\begin{eqnarray}
\label{eq:EQP}
\text{EQP}_\alpha = \frac{\sum_{l=1}^M \frac{(\beta_l - \widehat{\beta}_{\alpha l})^2}{\beta_l}}{1 + \sum_{k=1}^K (L_k - 1)},
\end{eqnarray}
where $\alpha$ indicates the correction method.
\section{Results}

\begin{figure}[H]
    \centering
    \includegraphics[width=1\linewidth]{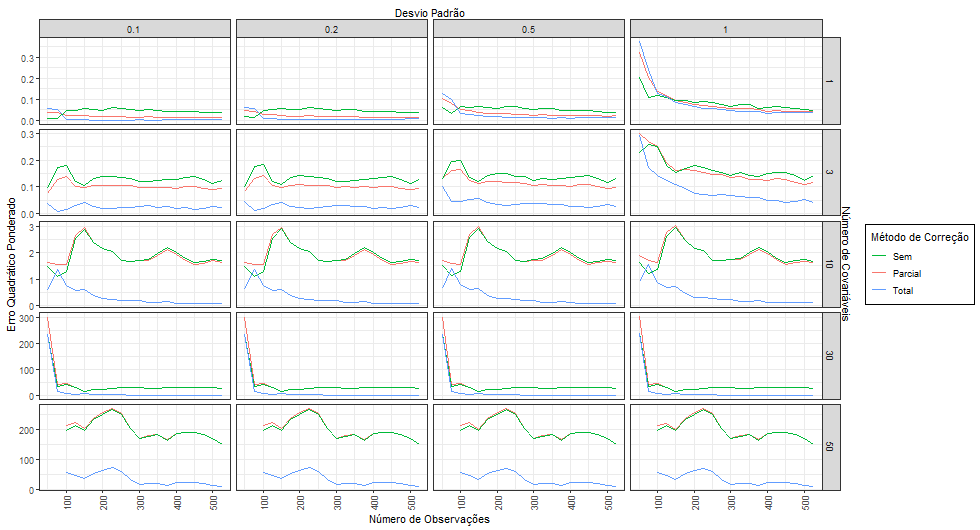}
    \caption{EQP calculado para o caso de pouca distorção com $L_k$ aleatório, para diferentes número de observações, número de variáveis categóricas e desvios padrões}
    \label{fig:resalto}
\end{figure}
\begin{figure}[H]
    \centering
    \includegraphics[width=1\linewidth]{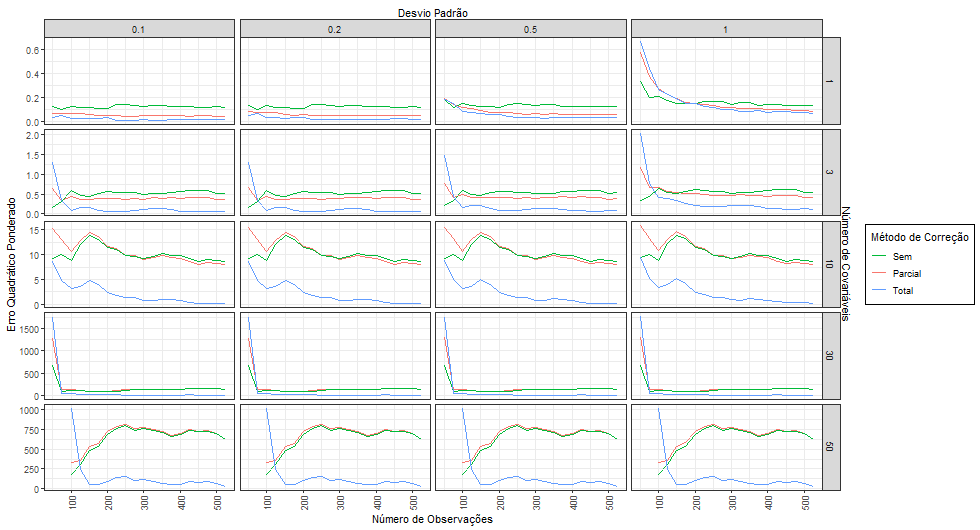}
    \caption{EQP calculado para o caso de média distorção com $L_k$ aleatório, para diferentes número de observações, número de variáveis categóricas e desvios padrões}
    \label{fig:resmedio}
\end{figure}
\begin{figure}[H]
    \centering
    \includegraphics[width=1\linewidth]{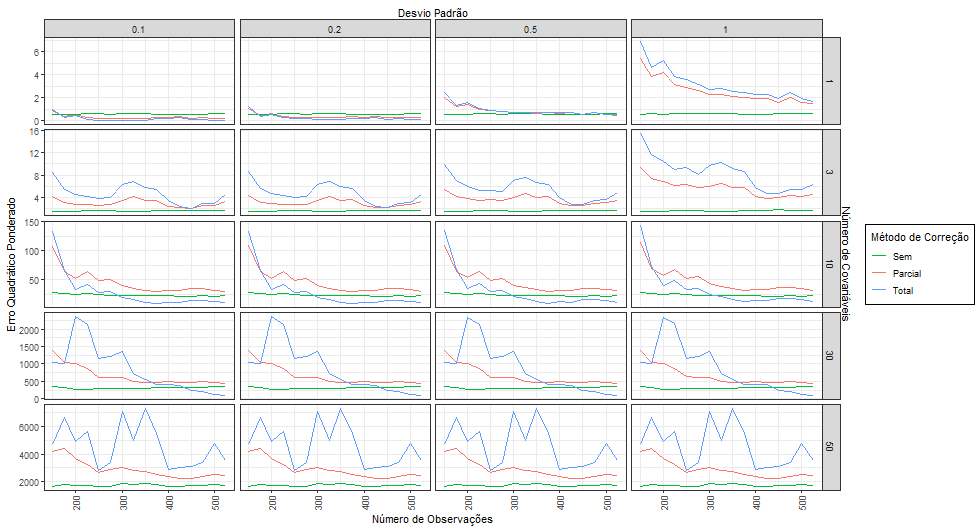}
    \caption{EQP calculado para o caso de alta distorção com $L_k$ aleatório, para diferentes número de observações, número de variáveis categóricas e desvios padrões}
    \label{fig:resbaixo}
\end{figure}
\begin{figure}[H]
    \centering
    \includegraphics[width=1\linewidth]{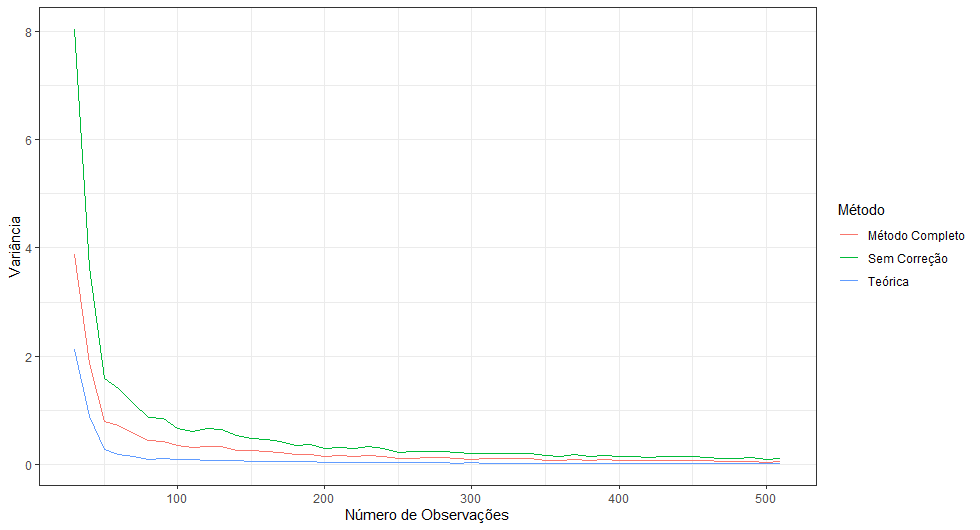}
    \caption{Variância do estimador do intercepto calculada para o caso de pouca distorção com $L_k=3, k=1,\ldots,K$ para os métodos de correção completa e sem correção, e a variância teórica do estimador}
    \label{fig:varianciateor}
\end{figure}
We simulated a scenario in which each categorical variable was randomly assigned a number of possible categories, with $L_k = 2$, $L_k = 3$, or $L_k = 4$ each occurring with equal probability $\frac{1}{3}$. From Figures~\ref{fig:resalto} and~\ref{fig:resmedio}, we observe that when $L_k$ is random, the difference between the full correction method and no correction becomes even more evident, with the corrected method outperforming in all cases except when there are small sample sizes with a high number of covariates, or a low number of covariates combined with a high standard deviation. Once again, from Figure~\ref{fig:resbaixo}, we can see that the correction method performs poorly under high distortion; however, in this case, when sample sizes are large and the number of covariates is moderate, applying the correction method is preferable to not correcting at all.

Furthermore, in Figure \ref{fig:varianciateor}, the variance values of the intercept estimator are compared between the full correction method and the uncorrected method, along with the theoretical variance described in Section~2.5. It is observed that the theoretical variance underestimates the variance actually observed, regardless of sample size. However, due to the asymptotic nature of the method, the theoretical variance approaches the observed variance as the sample size increases.

\section{Conclusion}
In this work, we propose an extension of the correction model developed by [Buonaccorsi et al., 2005], where we expand their ideas to a multinomial model. The problem described by that author lies in the fact that the observed covariates $\W$ may have classification errors, making them different from the true covariates $\X$. Specifically, our proposed correction uses the least squares estimators obtained through regression on 
$\W$, the marginal probabilities of 
$\X$, and the conditional probabilities of 
$\W$ given 
$\X$. In this way, it is possible to obtain corrected estimators without the need to observe the true values.

Due to the singularity of the covariance matrix of multinomial variables, it was not possible to jointly correct the intercept of the regression model along with the other coefficients. To overcome this limitation, we developed a method which, based on the corrected estimators, allows the intercept to be corrected. Simulation studies have shown that the proposed correction for the intercept is crucial for improving estimation.

Some assumptions were necessary for the calculation of these corrected estimators, particularly the independence between covariates and individuals. Although these assumptions are commonly used in linear models, they are not necessarily reasonable—especially in the field of genetics, where this correction could be particularly useful. Therefore, it is of utmost importance that future studies adapt the correction method by relaxing these assumptions.

Furthermore, the simulation studies conducted—especially in the high distortion case—could be improved by increasing the number of observations. Considering that the method is asymptotic, it is possible that a sufficiently large sample would yield better performance of the corrected estimator.


\backmatter


\section*{Acknowledgements}

This work was partially supported by FAPESP grants 2017/15306-9 and 2023/00592-7, and CNPq grant 444720/2024-3 and CAPES grant 001.



%


\appendix


\section{}
\subsection{Title of appendix}

\label{lastpage}

\end{document}